\documentclass[
aps,%
12pt,%
final,%
notitlepage,%
oneside,%
onecolumn,%
nobibnotes,%
nofootinbib,%
superscriptaddress,%
showpacs,%
centertags%
amsmath,%
amsfonts]%
{revtex4}

\def\'#1{{\accent19\ifx #1i \i\else #1\fi}}

\newcommand{\be}{\begin{equation}}
\newcommand{\ee}{\end{equation}}
\newcommand{\bea}{\begin{eqnarray}}
\newcommand{\eea}{\end{eqnarray}}
\newcommand{\beq}{\begin{equation}}
\newcommand{\eeq}{\end{equation}}
\newcommand{\beqa}{\begin{eqnarray}}
\newcommand{\eeqa}{\end{eqnarray}}
\newcommand{\noi}{\noindent}
\newcommand{\e}{\varepsilon}
\newcommand{\g}{{\mathfrak g}}

\newcommand{\f}{{\frac12}}
\newcommand{\iso}{{{\mathfrak{iso}}(1,3)}}
\newcommand{\so}{{{\mathfrak{so}}(1,3)}}
\newcommand{\s}{{\mathfrak{sl}(2)}}
\newcommand{\D}{{\cal D}}
\newcommand{\Sy}{{\cal S}}
\def\ZZ{{\mathbb Z}}

\newtheorem{theorem}{Theorem}[section]

\newtheorem{definition}[theorem]{Definition}
\newtheorem{example}[theorem]{Example}

\begin{document}
\title{Cubic extentions of the Poincar\'e algebra}
\author{\firstname{ Michel}~\surname{Rausch de Traubenberg}}
\email{rausch@lpt1.u-strasbg.fr}
\affiliation{Laboratoire de Physique Th\'eorique, CNRS UMR  7085,
  3 rue de l'Universit\'e, 67084 Strasbourg Cedex, France }

\begin{abstract}
A systematic study of  non-trivial cubic extensions of the four-dimensional 
 Poincar\'e algebra is undertaken.
Explicit examples are given with  various techniques (Young tableau,
characters {\it etc}).
\end{abstract}
 \maketitle

\section{Introduction}

The concept of symmetries has always been an essential tool in the
comprehension/description  of physical laws. 
Mathematically symmetries
are described by algebras. 
Of course,
not all the mathematical structures are relevant in physics since 
they have to respect the principles of relativity and quantum
mechanics. These consistency conditions 
lead to a series of no-go theorem \cite{cm, hls}, and
two wide classes of algebras namely, Lie  algebras
and Lie superalgebras were intensively used. 
However, these two structures
are not the only ones one is able to consider without contradicting
these  no-go theorems. 
It is then  appealing to know whether or
not new types of symmetry would be relevant in physics.
Several approaches beyond Lie (super)algebras have been considered.
The main idea of these various approaches is to  weaken
the hypotheses of the no-go theorems, obtaining new symmetries 
which go beyond supersymmetry  
\cite{ker, luis, para, kibler, FSUSY}.

In supersymmetric theories, the extensions of the Poincar\'e
algebra are obtained from a ``square root'' of the translations,
``$QQ \sim P$''.
It is tempting to consider other alternatives where the new algebra
is obtained from yet higher order roots.
The simplest alternative which we will consider in this paper is 
``$QQQ \sim P$''. It is important to stress that such structures
are not Lie (super)algebras  (even though they  contain a Lie sub-algebra),
and as such escape {\it a priori} the Coleman-Mandula \cite{cm}
as well as  the Haag-Lopuszanski-Sohnius no-go theorems \cite{hls}.
Furthermore, as far as we know, no no-go theorem associated to such types
of extensions has been considered in the literature. 
In such  extensions,
 the generators of the Poincar\'e algebra are obtained as   three-fold 
symmetric products of more fundamental generators, leading to the
``cubic root'' of translation.
The Lie algebras of order three, the structures which underlie these
extensions,
 were  defined in \cite{flie1, flie2} 
 in full analogy with supersymmetry and its underlying  Lie superalgebra structure.
An $F-$Lie algebra admits a $\mathbb Z_F-$gradation
($F=3$ in this paper), the zero-graded part
being a Lie algebra. An $F-$fold symmetric product (playing the role of
the anticommutator in the case $F=2$)  expresses the zero graded  part
in terms of the non-zero graded part 
\cite{flie1,flie2}.

Subsequently, a  specific $F-$Lie algebra (for $F=3$) has been studied 
together with its implementation in quantum field theory  \cite{mmr, mrt}. 
A general study of the possible non-trivial extensions of the
$(1+3)D$ Poincar\'e algebra has been undertaken. In this paper we
summarize some of the results established in \cite{grt}  and we give
a systematic way to construct all possible extensions of the Poincar\'e
algebra when $F=3$. 

The content of this paper is the following. In section 2. we recall the main
results on $F-$Lie algebras. Section 3. is devoted to the general study
of Lie algebras of order 3 associated to the Poincar\'e algebra.
Then, several examples are explicitly constructed using various techniques
such as Young tableau, characters {\it etc.}

\section{Lie algebras of order $F$}
The general definition of  Lie algebras of order $F$, was given in \cite{flie1, flie2} 
together with an inductive way to construct Lie algebras of order $F$ associated 
with {\it any} Lie algebra or Lie superalgebra.  We recall here the main results 
useful for the sequel.
Let  $F$  be a positive integer and define 
$q=e^{\frac{ 2 \pi i}{F}}$.  We consider $\g$ a complex vector space and $\e$ an automorphism
of $\g$ satisfying $\e^F=1$. Set $\g_i \subseteq \g$ the eigenspace corresponding to
the eigenvalue $q^i$ of $\e$. 
Then, we have $\g=\g_0 \oplus \cdots  \oplus \g_{F-1}$.

\begin{definition}
\label{flie}
The vector space  $\g = \g_0 \oplus \cdots \oplus \g_{F_1}$ is called a (complex)  Lie 
algebra of order $F$  if it is
endowed with the following structure:

\begin{enumerate}
\item $\g$ is a (complex) Lie algebra;
\item  
$\g_k, 1 \le k \le F-1$ are 
representations
of $\g$; 
\item there exist multilinear $\g-$equivariant maps 
$\{ \cdots \}: {\cal S}^F\left(\g_k\right)
\rightarrow \g_0$, where  ${ \cal S}^F(D)$ denotes
the $F-$fold symmetric product
of $D$;
\item
for all  $Y_1,\cdots,Y_{F+1} \in \g_k$ the following
``Jacobi identities'' hold:

\beqa
\label{J4}
\sum\limits_{i=1}^{F+1} \left[ Y_i,\left\{Y_1,\dots,
Y_{i-1},
Y_{i+1},\dots,Y_{F+1}\right\} \right] =0.
\eeqa
\end{enumerate}
\end{definition}

\noindent
It should be noted that
if $F=1$, by definition $\mathfrak{g}=\mathfrak{g}_0$ and a Lie algebra of order $1$ is 
a Lie algebra; and   if $F=2$, then $\g$ is a Lie superalgebra. Therefore,  
Lie algebras of order $F$ appear as some  kind of generalisations of Lie algebras and 
Lie superalgebras. Moreover, for any $k=1,\ldots,F-1$, the 
$\ZZ_F-$graded vector spaces ${\mathfrak{g}}_0\oplus{\mathfrak{g}}_k$ 
is  a Lie algebra of order $F$. We  call these type of algebras 
\textit{elementary Lie algebras of order $F$}. 
From now on we consider only elementary Lie
algebras of order 3.

In \cite{flie2} an inductive process for the construction of Lie algebras of order $F$ starting 
from a Lie algebra of order $F_1$ with
$1\le F_1$ was proved

\begin{theorem} \label{tensor} 

\noi
Let $\mathfrak{g_0}$ be a Lie algebra and $\mathfrak{g}_1$ be
$\mathfrak{g_0}-$module  such that    

(i) $\g=\mathfrak{g_0} \oplus \mathfrak{g_1}$ is a Lie algebra of
order $F_1>1$; 

(ii) $\mathfrak{g_1}$ admits  a $\mathfrak{g_0}-$equivariant symmetric form
 of order $F_2 > 1$.

\noindent
Then  $\g=\mathfrak{g_0}  \oplus \mathfrak{g_1}$ admits 
a Lie algebra of order $F_1 +F_2$ structure.
\end{theorem}

\noi
 We give now two examples of Lie algebras of order 3 obtained
through this construction.

 \begin{example}
\label{so23}
Let $\g_0$ be a semi-simple Lie algebra and $\g_1$ its adjoint representation.  
Let $\{J_a, a =1,\cdots, \text{dim}\  \g_0\}$
 be a basis of $\g_0$ 
and $\{A_a, a =1,\cdots, \text{dim}\  \g_0\}$ be the
 corresponding basis of $\g_1$. 
Let $g_{ab}=Tr(A_aA_b)$ be the Killing form and $f_{ab}{}^c$ be the structure 
constants of $\g_0$. 
Then one can endow $\g=\g_0\oplus\g_1$ with a 
Lie algebra of order $3$ structure 
\beq
[J_a,J_b]=f_{ab}{}^c J_c, \ \ [J_a,A_b]=f_{ab}{}^c A_c, \ \ 
\{ A_a,A_b,A_c \}=g_{ab}J_c+g_{ac}J_b+g_{bc}J_a. \nonumber
\eeq
\end{example}

\noi
(This example is a consequence of the Theorem above, modified
to include $F_1=1.$)

\begin{example}
\label{FP}
Let $\g_0 = \left< L_{\mu \nu }, P_\mu\right>$ be the $D-$dimensional Poincar\'e algebra and  
$\g_1=\left<V_\mu \right>$ be the $D-$dimensional vector representation of $\g_0$.
The brackets
\beqa
 \left[L_{\mu \nu }, L_{\rho \sigma}\right]&=&
\eta_{\nu \sigma } L_{\rho \mu }-\eta_{\mu \sigma} L_{\rho \nu} + \eta_{\nu \rho}L_{\mu \sigma}
-\eta_{\mu \rho} L_{\nu \sigma},\nonumber \\
\left[L_{\mu \nu }, P_\rho \right]&=& \eta_{\nu \rho } P_\mu -\eta_{\mu \rho } P_\nu,  \ 
\left[L_{\mu \nu }, V_\rho \right]= \eta_{\nu \rho } V_\mu -\eta_{\mu \rho } V_\nu, \
\left[P_{\mu}, V_\nu \right]= 0, \nonumber \\ 
\{ V_\mu, V_\nu, V_\rho \}&=&
\eta_{\mu \nu } P_\rho +  \eta_{\mu \rho } P_\nu + \eta_{\rho \nu } P_\mu,
\nonumber
\eeqa
\noi
with the metric $\eta_{\mu \nu}=\rm{diag}(1,-1,\cdots,-1)$ endow 
$\g=\g_0 \oplus \g_1$ with 
an elementary Lie algebra of order $3$ structure.
\end{example}
It has been shown that  Example \ref{so23} (with $\g_0=\mathfrak{so}(2,3)$) and Example \ref{FP}
(when $D=4$) are
related through an In\"on\"u-Wigner contraction \cite{flie2}.
The algebra of Example  \ref{FP}, firstly introduced in \cite{flie2}, has
been studied in \cite{mmr} together with its implementation in 
Quantum Field Theory when  $D=4$. Subsequently it has been realised
\cite{pform} that in arbitrary dimension this algebra acts  in
a natural geometric way on generalised gauge field or $p-$forms.

\section{Extension of the Poincar\'e algebra}
The $(1+3)-$dimensional Poincar\'e algebra $\iso$ is given by
\beqa
 \left[L_{\mu \nu }, L_{\rho \sigma}\right]=
\eta_{\nu \sigma } L_{\rho \mu }-\eta_{\mu \sigma} L_{\rho \nu} + \eta_{\nu \rho}L_{\mu \sigma}
-\eta_{\mu \rho} L_{\nu \sigma}, \ \
\left[L_{\mu \nu }, P_\rho \right]= \eta_{\nu \rho } P_\mu -\eta_{\mu \rho } P_\nu.
\eeqa
As it is well known the following change of basis (precisely in the complexified
of the Lorentz algebra) 
$ L_{\mu \nu} \to (J_i= \frac12 \varepsilon_{i}{}^{jk} L_{jk}, K_i=L_{0i}) \to 
(N_i=\frac12(J_i +i K_i), \bar N_i=\frac12(J_i -i K_i))$
leads to 
$[N_i,N_j]=\varepsilon_{ij}{}^kN_k,\ [\bar N_i,\bar N_j]=\varepsilon_{ij}{}^k\bar N_k,
\  [N_i,\bar N_j]=0$ 
(with $\varepsilon_{ij}{}^k=\varepsilon_i{}^{jk}=\e_{ijk}$ the Levi-Civita tensor).
Therefore, a finite dimensional irreducible representation of $\mathfrak{so}(1,3)$
is specified by the eigenvalues of the two Casimir operators $Q=N_1^2 + N_2^2 +N_3^2$
and $\bar Q=\bar N_1^2 + \bar N_2^2 +\bar N_3^2$ which are 
$a(a+1)$ and $b(b+1)$ respectively.
We denote $\D_{a,b}, \ a, b \in \frac12 \mathbb N$ the corresponding representation of dimension
$(2a+1)(2b+1)$.\\

Now, from the Poincar\'e algebra $\iso$ and a given (reducible)
representation $\g_1$
of $\so$ we construct non-trivial extensions of the Poincar\'e algebra
in three steps:
\begin{enumerate}
\item we extend the action of $\so$ on $\g_1$ to the action of
$\iso$ on $\g_1$;
\item we study all possible $\so-$equivariant mappings from
$\Sy^3(\g_1) \to \D_{\frac12,\frac12}$;
\item we impose the Jacoby identities which ensure the consistency of
 the algebra.
\end{enumerate}

\subsection{Finite dimensional representations of the Poincar\'e algebra}
Let $\g_1=  \oplus_i \D_{a_i,b_i} $ be an arbitrary reducible finite dimensional 
representation of $\so$.
We would like to extend the action of 
$\mathfrak{so}(1,3)$
on $\g_1$, to the action of the   Poincar\'e algebra  on
$\g_1$.  Namely, we would like to calculate $[P_\mu, \D_{a,b}]$.
Usually, in field theory,
starting from a finite dimensional 
representation $\g_1$ of $\so$, a non-trivial representation
of $\iso$ is realised in terms of an infinite dimensional representation, the field,
with $P_\mu = \partial_\mu$. Here, we would like to see whether or not $P_\mu$ 
may act non-trivially on $\g_1$ {\it i.e.} can be represented by   finite dimensional matrices.

If  $\g_1$ is an irreducible representation then 
one can show that $P_\mu$ acts trivially on $\g_1$,
{\it i. e.} $[P_\mu,\g_1]=0$ \cite{grt}.
The case where $\g_1$ is reducible is more involved as can be seen of the 
following two examples:

\begin{itemize}
\item[(1)] If $\g_1 = {\cal D}_{\frac12,\frac12} \oplus {\cal D}_{0.0}$ the vector plus the
scalar representations of
$\so $,  $P_\mu$ can be represented by the $5 \times 5$ nilpotent
matrices

$$P_\mu=\left(\begin{array}{ccccc}
0&0&0&0&\delta_{\mu}^0 \\
0&0&0&0&\delta_{\mu }^1 \\
0&0&0&0&\delta_{\mu }^2 \\
0&0&0&0&\delta_{\mu }^3 \\
0&0&0&0&0 \end{array}\right). 
$$

Indeed, if we denote $\left<v_\mu, \mu=0,\cdots,3\right>$
(resp. $\left<w_0\right>$) a basis of ${\cal D}_{\frac12,\frac12}$ 
(resp.  $ {\cal D}_{0,0}$) we have
$ P_\mu (w_0)= v_\mu, \ \ P_\mu (v_\nu) =0.$

\item[(2)] If $\g_1 = {\cal D}_{\frac12,0} \oplus {\cal D}_{0,\frac12}$,
 since $\D_{0,\frac12} \otimes \D_{\frac12,\frac12} = \D_{\frac12,1} \oplus \D_{\frac12,0} 
\supset \D_{\frac12,0}$,  
$P_\mu$ can be represented by the  $4 \times 4$ nilpotent  matrices
$P_\mu=\left(\begin{array}{cc} 
0&\sigma_\mu \\0&0
\end{array}\right)$

(with $\sigma_0$ the $2 \times 2$ identity matrix and $\sigma_i, i=1,2,3$ 
the Pauli matrices)
such that for $\psi \in \D_{\frac12,0}, \bar \chi \in \D_{0,\frac12}$ 
 we have
$ P_\mu (\psi)=0, \ \ P_\mu( \bar \chi) = \sigma_\mu \bar \chi \in \D_{\f,0}.$
\end{itemize}

The general case can be more complicated and his synthesize  in Lemma 5.1 of \cite{grt}.
Here, we just recall the main properties of this technical Lemma. Let $\g_1$ be a 
finite-dimensional (reducible)
representation of $\so$, such that the space-time translations act non-trivially on $\g_1$.

Then, 

\begin{enumerate}
\item $P_\mu$ are represented by nilpotent matrices;
\item  if $\D_{a,b}$ and $D_{c,d}$ are two irreducible representations 
such that 
$P_\mu : \D_{a,b} \longrightarrow \D_{c,d}$ non-trivially, then 
$\D_{c,d} \subseteq \D_{a,b} \otimes \D_{\f,\f}$;
\item the representation $\g_1$ is indecomposable.
\end{enumerate} 

\subsection{$\so-$equivariant mappings}
\subsubsection{Construction of $\so-$equivariant mappings}
Next, we
construct the possible $\so-$equivariant mappings
 from $\Sy^3(\g_1)$ into
$\D_{\frac12,\frac12}$, with $\g_1$ an arbitrary representation of
$\so$.
We recall the following  isomorphisms of representations of
$GL(A) \times GL(B)$ \cite{fulton-harris} (p. 80):

\beqa
\label{sum}
{\cal S}^3 \left(A \oplus B \oplus C\right)&\cong&
{\cal S}^3 \left( A\right) \ \oplus  \ {\cal S}^3 \left( B \right)
\ \oplus \  {\cal S}^3 \left( C\right) 
\ \oplus {\cal S}^2 \left( A\right) \otimes B  
 \ \oplus \
{\cal S}^2 \left( A\right) \otimes C
 \ \oplus \ {\cal S}^2 \left( B\right) \otimes A
\nonumber \\
  &\oplus&
{\cal S}^2 \left( B\right) \otimes C \ \oplus \
{\cal S}^2 \left( C\right) \otimes A \ \oplus \
{\cal S}^2 \left( C\right) \otimes B  
\ \oplus\  A \otimes B \otimes C, 
 \nonumber \\
{\cal S}^3 \left(A \otimes B\right) &\cong& 
{\cal S}^3 \left(A\right) \otimes {\cal S}^3 \left(B\right)\  \oplus \
{\$}^{^{{\tiny \begin{tabular}{|c|c|}\hline
\ \  & \ \  \\ \hline
\  \\
\cline{1-1}
\end{tabular}}}} \hskip -.3truecm
\left(A\right) \otimes
{\$}^{^{{\tiny \begin{tabular}{|c|c|}\hline
\ \ & \ \  \\ \hline
 \\
\cline{1-1}  
\end{tabular}}}} \hskip -.3truecm \left(B\right) \ \oplus \
\Lambda^3\left(A\right) \otimes \Lambda^3\left(B\right),  \\
{\cal S}^2 \left(A \otimes B\right) &\cong& 
{\cal S}^2 \left(A\right) \otimes {\cal S}^2 \left(B\right)\  \oplus \
\Lambda^2\left(A\right) \otimes \Lambda^2\left(B\right), \nonumber 
\eeqa

\noindent
where $\Sy^p(D)$ (resp. $\Lambda^p(D), 
{\$}^{^{{ {{\hbox{{\tiny {\tiny\renewcommand
\arraycolsep{0.1pt} 
\begin{tabular}{|c|c|}\hline
\ \ & \ \  \\ \hline
 \\
\cline{1-1}  
\end{tabular}}}}}}}}} \hskip -.3truecm \left(D\right) $) denotes 
the irreducible
representations $GL(A)$ symmetric (resp. antisymmetric,  corresponding to the Young 
symmetriser of the Young diagram {\tiny $  
\begin{tabular}{|c|c|}\hline 
\ \ & \ \  \\ \hline 
 \\ 
\cline{1-1}   
\end{tabular}$}).

Let $\g_1$ be a representation of $\so$
and let $\D_{\frac12,\frac12}$ be the vector representation of $\so$.
Using the first equation given in (\ref{sum}), since $\g_1$ is a reducible
representation of $\so$, $\ \Sy^3(\g_1)$ reduces to three types
of terms (i) ${\cal S}^3 \left({\cal} \D\right)$, 
(ii) ${\cal S}^2 \left({\cal} \D\right)  \otimes \D' $
and (iii) ${\cal D} \otimes {\cal D}'\otimes  {\cal D}''$
with ${\cal D}, {\cal D}',{\cal D}''$ three irreducible representations.
Thus all possible $\so-$equivariant mappings are of the type
(i) ${\cal S}^3 \left({\cal} \D\right) \longrightarrow {\cal D}_{\f,\f}$, 
(ii) ${\cal S}^2 \left({\cal} \D\right)  \otimes \D' 
\longrightarrow {\cal D}_{\f,\f}$  and 
(iii) ${\cal D} \otimes {\cal D}'\otimes  {\cal D}''
\longrightarrow {\cal D}_{\f,\f}$. We now characterise more precisely these
mappings. (From now one, $\D_{a,b}$ is written 
$\D_{a,b} = \D_{a,0} \otimes \D_{0,b}$   with
$[Q,\D_{a,0}]=a(a+1)\D_{a,0},\ [Q,\D_{0,b}]=0, \
[\bar Q,\D_{a,0}]=0,\ [\bar Q,\D_{0,b}]=b(b+1)\D_{0,b}$
where $Q, \bar Q$ are the two Casimir operators of $\so$.)
\\

\medskip
\noindent
(i) \underline{Type $I$. $\so-$ equivariant mappings}:
${\cal S}^3 \left({\cal} \D\right) \longrightarrow {\cal D}_{\frac12,\frac12}$
\\

\noi
Let ${\cal D} =\D_{a,b}= {\cal D}_{a,0} \otimes {\cal D}_{0,b}$ with $a,b \in
\frac12 \mathbb N$, 
${\cal D}_{\frac12,\frac12} \subseteq {\cal D}_{a,b}\otimes {\cal D}_{a,b}\otimes 
{\cal D}_{a,b}$ if $a$ and $b$ are half-integer.
From the second  equation of (\ref{sum})
${\cal D}_{\frac12,\frac12} \subseteq {\cal S}^3\left({\cal D}_{a,b}\right)$ if
either 
\begin{enumerate}
\item[$I_S$ :] ${\cal D}_{\frac12,0} \subseteq {\cal S}^3\left({\cal D}_{a,0}\right)$ and
${\cal D}_{0,\frac12} \subseteq {\cal S}^3\left({\cal D}_{0,b}\right)$;
\item[$I_A$ :]  ${\cal D}_{\frac12,0} \subseteq \Lambda^3\left({\cal D}_{a,0}\right)$ and
${\cal D}_{0,\frac12} \subseteq \Lambda^3\left({\cal D}_{0,b}\right)$;
\item[$I_M$ :]
${\cal D}_{\frac12,0} \subseteq {\$}^{^{{\tiny \begin{tabular}{|c|c|}\hline
\ \ & \ \  \\ \hline
 \\
\cline{1-1}  
\end{tabular}}}} \hskip -.3truecm \left({\cal D}_{a,0}\right)$ and
${\cal D}_{0,\frac12} \subseteq {\$}^{^{{\tiny \begin{tabular}{|c|c|}\hline
\ \ & \ \  \\ \hline
 \\
\cline{1-1}  
\end{tabular}}}} \hskip -.3truecm \left({\cal D}_{0,b}\right)$.
\end{enumerate} 
In particular, when   ${\cal D}={\cal D}_{a,a}$ and  $a$ half-integer,
the mapping ${\cal S}^3\left({\cal D}_{a,a}\right)\longrightarrow
{\cal D}_{\frac12,\frac12}$
is {\it always} $\so-$ equivariant and is called
type $I_0{}_S,I_0{}_A,I_0{}_M$ respectively. 
The extension of the Poincar\'e algebra given in 
Example \ref{FP} is of  type  $I_0{}_M$ with $a=b=\f$.\\

\noi
(ii) \underline{Type $II$. $\so-$ equivariant mappings}:
 ${\cal S}^2 \left({\cal} \D\right)  \otimes \D' 
\longrightarrow {\cal D}_{\frac12,\frac12}$ 

\noi
Let ${\cal D}= {\cal D}_{a,b}$ and ${\cal D}'= {\cal D}_{c,d}$,
${\cal D}_{\frac12,\frac12} \subseteq {\cal D}_{a,b}\otimes 
{\cal D}_{a,b}\otimes 
{\cal D}_{c,d}$ if  $c,d$ are half-integer
and there exists an $n=0,\cdots,2a$ such that $ 2a-n-b=\frac12$ or $b-2a + n =\frac12$.
 From the third   equation of (\ref{sum})
${\cal D}_{\frac12,\frac12} \subseteq {\cal S}^2\left({\cal D}_{a,b}\right)
\otimes {\cal D}_{c,d}$ if either 

\begin{enumerate}
\item[$II_S$ :] ${\cal D}_{\frac12,0} \subseteq {\cal S}^2\left({\cal D}_{a,0}\right)
\otimes {\cal D}_{c,0}$ and
${\cal D}_{0,\frac12} \subseteq {\cal S}^2\left({\cal D}_{0,b}\right)
\otimes {\cal D}_{0,d}$;
\item[$II_A$ :]  ${\cal D}_{\frac12,0} \subseteq \Lambda^2\left({\cal D}_{a,0}\right)
\otimes {\cal D}_{c,0}$ and
${\cal D}_{0,\frac12} \subseteq \Lambda^2\left({\cal D}_{0,b}\right)
\otimes {\cal D}_{0,d}$.
\end{enumerate}
We call these  
 $\so-$ equivariant 
mappings,  mappings of type 
$II_S$ and  $II_A$ 
 respectively. \\

\noi
(iii) \underline{Type $III$. $\so-$ equivariant mappings}:
 ${\cal D} \otimes {\cal D}'\otimes  {\cal D}''
\longrightarrow {\cal D}_{\frac12,\frac12}$. 
\\

\noi
Let ${\cal D}= {\cal D}_{a,b}, {\cal D}'= {\cal D}_{c,d}$
and ${\cal D}''= {\cal D}_{e,f}$  ($ a \ge c \ge e$).  If 
 $(a+c+e)$ is half-integer and if there exists an
$n= 0, \cdots, 2c$ such that $a+c-n-e=\frac12$ or $e-a-c+n=\frac12$
(plus similar relations for $b,d,f$) then 
${\cal D}_{\f,\f} \subseteq {\cal D}_{a,b}\otimes {\cal D}_{c,d}\otimes 
{\cal D}_{e,f}$. Thus,
there are many $\so-$equivariant mappings of these
types. \\ 

\subsubsection{Explicit construction using Young projectors}

In this subsection, we give an explicit $\so-$equivariant mapping 
by mean of Young projectors.
 Let $\D=\D_{\f,0}\otimes \D_{0,\f} $.
Using conventional notations for spinors,
let $\D_{\f,0} = \left<\psi_\alpha, \alpha=1,2\right>$
and $\D_{0,\f}=  \left<\bar \chi^{\dot \alpha}, \dot \alpha=1,2\right>$
be the spinor representations of $\so$. We 
introduce the Dirac $\Gamma-$matrices


$$\Gamma_\mu =\left(\begin{array}{cc}
 0& \sigma_\mu \\ \bar \sigma_\mu&0
\end{array}\right),$$

\noi
with $\sigma_\mu=(\sigma_0,\sigma_i)$ and $\bar \sigma_\mu=(\sigma_0,-\sigma_i)$.
 The index structure of the $\sigma-$matrices
is as follow $\sigma_\mu \to \sigma_\mu{}_{\alpha \dot \alpha},
\bar \sigma_\mu \to \bar \sigma_\mu{}^{\dot \alpha  \alpha}$.
We also define 
$\psi_\alpha =\varepsilon_{\alpha\beta}\psi^\beta$,
$\psi^\alpha =\varepsilon^{\alpha\beta}\psi_\beta$,
$\bar\chi_{\dot\alpha}=\bar \varepsilon_{\dot\alpha
\dot\beta}\bar\chi^{\dot\beta}$, $\bar\chi^{\dot\alpha}
=\bar \varepsilon^{\dot\alpha\dot\beta}\bar\chi_{\dot\beta}$
with the antisymmetric  matrices 
$\varepsilon, \bar \varepsilon$  given by
$\varepsilon_{12} = \bar \varepsilon_{\dot 1\dot 2}=-1$,
$\varepsilon^{12} = \bar \varepsilon^{\dot 1\dot 2}=1$.

Now,  we consider the representation

$$
\D'_{\f,0}\cong {\$}^{^{{\tiny \begin{tabular}{|c|c|}\hline
1& 3\\ \hline
2 \\
\cline{1-1}  
\end{tabular}}}} \hskip -.3truecm \left({\cal D}_{\f,0}\right).
$$

\noi
We introduce the projector  (Young symmetriser)\footnote{Usually the other
convention is taken for the Young symmetriser :
$P=\frac13(1+(13))(1-(12))$. But here it is more convenient
to choose (\ref{Pyoung}) for the calculus.}

\beqa
\label{Pyoung}
P_{_{{\tiny \begin{tabular}{|c|c|}\hline
1& 3\\ \hline
2 \\
\cline{1-1}  
\end{tabular}}}}= \frac13(1-(12))(1+(13))=\frac13( 1-(12) +(13) - (123))
\eeqa

\noi
with 

$$(a\  b) = \left(\begin{array}{cc}  a &b \\
                                b&a\end{array}\right), 
(a\   b \ c ) = \left(\begin{array}{ccc} a &b & c\\
                                b&c&a\end{array}\right)$$

\noi two
cycles of ${\cal S}_3$ the group of permutation with three elements.
A direct calculation gives 

\beqa
\label{spin}
 P_{_{{\tiny \begin{tabular}{|c|c|}\hline
1& 3\\ \hline 
2 \\ 
\cline{1-1}   
\end{tabular}}}} \Big(\psi_{\alpha} \otimes \psi_{\beta} \otimes 
\psi_{\gamma}\Big)
=\varepsilon_{\alpha \beta} \lambda_\gamma
\eeqa

\noi
with $\D'_{\f,0}=\left<\lambda_\alpha, \alpha=1,2\right>$,
$\lambda_a= \Big(\psi_1\otimes \psi_2-\psi_2 \otimes \psi_1\Big)\otimes \psi_a
-\psi_a\otimes\Big(\psi_1\otimes \psi_2-\psi_2 \otimes \psi_1\Big)$
(the same result can be obtained using the usual calculus of the
Clebsch-Gordan coefficients).
Proceeding along the same lines with $\D_{0,\f}$ and
introducing 

$$ \D'_{0,\f}\cong {\$}^{^{{\tiny \begin{tabular}{|c|c|}\hline
1& 3\\ \hline
2 \\
\cline{1-1}  
\end{tabular}}}} \hskip -.3truecm \left({\cal D}_{0,\f}\right)
=\left<\bar \rho^{\dot \alpha} ,\dot \alpha=1,2\right>$$

\noi
we obtain

$$ 
 P_{_{{\tiny \begin{tabular}{|c|c|}\hline
1& 3\\ \hline 
2 \\ 
\cline{1-1}   
\end{tabular}}}} \Big(\psi_{\alpha} \otimes \psi_{\beta} \otimes 
\psi_{\gamma}\Big)
\bigotimes
 P_{_{{\tiny \begin{tabular}{|c|c|}\hline
1& 3\\ \hline 
2 \\ 
\cline{1-1}   
\end{tabular}}}} \Big(\bar \chi_{\dot \alpha} \otimes 
\bar \chi_{\dot \beta} \otimes 
\bar \chi_{\dot \gamma}\Big)=\varepsilon_{\alpha \beta} \bar 
\varepsilon_{\dot \alpha
\dot \beta} \lambda_\gamma \otimes \bar \rho_{\dot \gamma}.$$

\smallskip
\noi
Symmetrising the R.H.S. we then get

\beqa
\label{fpoincare}
{\cal S}^3\Big((\psi_\alpha \otimes \bar \chi_{\dot \alpha})
\otimes (\psi_\beta \otimes \bar \chi_{\dot \beta}) \otimes
 (\psi_\gamma \otimes \bar \chi_{\dot \gamma})\Big)=
\varepsilon_{\alpha \beta} \bar \varepsilon_{\dot \alpha \dot \beta} 
\lambda_\gamma \otimes \bar \rho_{\dot \gamma} +
\varepsilon_{\gamma \alpha } \bar \varepsilon_{\dot \gamma \dot \alpha} 
\lambda_\beta \otimes \bar \rho_{\dot \beta} +
\varepsilon_{\beta \gamma} \bar \varepsilon_{\dot \beta \dot \gamma} 
\lambda_\alpha \otimes \bar \rho_{\dot \alpha}. \nonumber \\
\eeqa
Now, from the isomorphism of $\D_{\f,0} \otimes \D_{0,\f}$
with the vector representation, 
 we have 
the correspondence 

\beqa
\label{spin-vec}
\begin{array}{ll}
V_\mu =\bar \sigma_\mu{}^{\dot \alpha \alpha} \psi_\alpha \otimes 
\bar \chi_{\dot \alpha}, & \psi_\alpha \otimes 
\bar \chi_{\dot \alpha} = \frac12 \sigma^\mu{}_{\alpha \dot \alpha} V_\mu,\\
P_\mu =\bar \sigma_\mu{}^{\dot \alpha \alpha} \lambda_\alpha \otimes 
\bar \rho_{\dot \alpha}, & \lambda_\alpha \otimes 
\bar \rho_{\dot \alpha} = \frac12 \sigma^\mu{_{\alpha \dot \alpha}} P_\mu,
\end{array}
\eeqa

\noi
(thus $\left<P_\mu, \ \mu=0,\cdots,3\right> \sim \D_{\f,\f}$,
 $\left<V_\mu, \ \mu=0,\cdots,3\right> \sim \D'_{\f,\f}$)
  and equations (\ref{fpoincare}) reduce to 

\beqa
\label{s3}
\Sy^3(V_\mu\otimes V_\nu \otimes V_\rho)=
\eta_{\mu \nu } P_\rho + \eta_{\nu \rho } P_\mu + \eta_{\rho \mu } P_\nu,
\eeqa

\noi
which is the trilinear bracket of the algebra given in Example \ref{FP}

\subsubsection{Decomposition of $\D\otimes \D \otimes \D$ with
respect to symmetry of Young diagrams}

The $\so-$ equivariant mappings of type $I$  lead immediately to an interesting
question. Let $\D_{a}$ be an irreducible representation of $\s$\footnote{
Since $\mathfrak{so}(1,3,\mathbb C)\sim \s \oplus \s$.}.
It is well known that 
$\D_a \otimes \D_a= \Sy^2(\D_a) \oplus \Lambda^2(\D_a)$
with $\Sy^2(\D_a) = \D_{2a} \oplus \D_{2a-2} \oplus \cdots$,
$\Lambda^2(\D_a)=\D_{2a-1} \oplus \D_{2a-3} \oplus \cdots$,
that is one can identified each irreducible summand of $\D_a \otimes \D_a$
in the symmetric or antisymmetric part of $\D_a \otimes \D_a$.
If we now consider  $\D_a \otimes \D_a \otimes \D_a$ similarly  we have
the following decomposition

$$\D_a\otimes \D_a \otimes \D_a = \Sy^3(\D_a) \oplus
{\$}^{^{{\tiny \begin{tabular}{|c|c|}\hline
 \ \ &  \ \ \\ \hline
 \\
\cline{1-1}  
\end{tabular}}}} \hskip -.3truecm \left({\cal D}_{a}\right)
\oplus \Lambda^3(\D_a)
$$

\noi and we would like to identify in which symmetry of the Young
tableau is a given irreducible summand. To answer to this
question, we firstly have to recall
some known results on the character.

Let $\g$ be a semisimple Lie algebra and denote $\Lambda$ the weight
lattice. The integral group ring ${\mathbb Z}[\Lambda]$ 
on the abelian group $\Lambda$ is
the free $\mathbb Z-$module $\mathbb Z^{(\Lambda)}$. If we write $e^\lambda$
the basis element of ${\mathbb Z}[\Lambda]$ corresponding to the weight
$\lambda$ we have the multiplication law $e^\lambda e^\mu = e^{\lambda + \mu}$.
If now, we introduce the representation ring $R(\g)$ the character 
homomorphism is defined\footnote{
This formal character can be identified with the usual character 
of the corresponding representation of the Lie group G (associated to the  
Lie algebra $\g$), restricted to the Cartan subgroup 
\cite{fulton-harris} (p. 381).}  
by

\beqa
\label{character}
\begin{array}{lrll} \text{ch}:& R(\g) &\to& {\mathbb Z}[\Lambda] \\
                               & V&\mapsto& \sum \text{dim} V_\lambda e^\lambda
            \end{array}
\eeqa

\noi where $V_\lambda$ is the weight space corresponding to the 
weight $\lambda$ of the representation space $V$ \cite{fulton-harris}
(p. 375).
The representation ring also has the structure of $\lambda-$ring, this
allows to express the character of the symmetric and exterior power
through the formul\ae
 
\beqa
\label{char}
\sum \limits_{n \in \mathbb N} \text{ch}\left(\Sy^n(V)\right) T^n
&=&\exp\left({\sum \limits_{m \ge 1} \frac{1}{m}\Psi^m(\text{ch}(V) )T^m}
\right), 
\nonumber \\
\sum \limits_{n \in \mathbb N} \text{ch}\left(\Lambda^n(V)\right) T^n
&=&\exp\left({\sum \limits_{m \ge 1} \frac{\ \ (-1)^{m-1}}{m}\Psi^m(\text{ch}
(V)) T^m}\right), 
\eeqa

\noi
where $\Psi^m$ is the Adams linear operator 
$\Psi^m(e^\lambda)= e^{m \lambda}$ \cite{bourbaki} (exercise 11, p. 231).
Developing both side of (\ref{char}) up to $T^3$ we get

\beqa
\label{sym}
\text{ch}(\Sy^2(V))&=&\frac12 \Psi^2(\text{ch} V) +
\frac12(\text{ch} V)^2 
\nonumber 
\\
\text{ch}(\Sy^3(V))&=&\frac13 \Psi^3(\text{ch} V) 
+\frac12 \text{ch} V
 \Psi^2(\text{ch} V)+
\frac16(\text{ch} V)^3  
\eeqa

\noi
and 

\beqa
\label{exterior}
\text{ch}(\Lambda^2(V))&=&-\frac12 \Psi^2(\text{ch} V) +
\frac12(\text{ch} V)^2 \nonumber \\
\text{ch}(\Lambda^3(V))&=&\frac13 \Psi^3(\text{ch} V) 
-\frac12 \text{ch} V 
 \Psi^2(\text{ch} V)+
\frac16(\text{ch} V)^3. 
\eeqa

\noi
Finally, since

$$\left(\text{ch} V\right)^3= \text{ch}(\Sy^3(V)) +
\text{ch}({\$}^{^{{\tiny \begin{tabular}{|c|c|}\hline
 \ \ &  \ \ \\ \hline
 \\
\cline{1-1}  
\end{tabular}}}} \hskip -.3truecm \left(V\right)+
\text{ch}(\Lambda^3(V)),$$
we get 

\beqa
\label{mixed}
\text{ch}({\$}^{^{{\tiny \begin{tabular}{|c|c|}\hline
 \ \ &  \ \ \\ \hline
 \\
\cline{1-1}  
\end{tabular}}}} \hskip -.3truecm \left(V\right))=
-\frac23 \Psi^3(\text{ch}V) +\frac23\left(\text{ch} V\right)^3.
\eeqa

\noi
For $\s$ the weight lattice in one dimensional,
and if we denote $(e^n)_{n \in {\mathbb Z}}$ a basis of the integral ring group
$\mathbb Z[\Lambda]$, for the representation $\D_a$ the character
(\ref{character}) reduces to

\beqa
\label{charsl2}
\text{ch}(\D_a)=e^{-a} + e^{-a+1} + \cdots e^{a-1} + e^a.
\eeqa

\noi
Let us mention that (\ref{charsl2}) also comes from the Weyl character
formula. Indeed, if we denote $\mu$ the positive weight of $\s$,
$\rho=\sum \limits_{\text{positive weight } \lambda} \frac{\lambda}{2}=
\frac{\mu}{2}$. For the representation $\D_a$ the highest weight is
$\lambda=a \mu$  and the Weyl character formula simplifies to

$$
\text{ch}(\D_a)=\frac{e^{a \mu + \rho}-e^{-(a \mu+\rho)}}
{e^{ \rho}-e^{-\rho}}$$ and  gives (\ref{charsl2})
(with $e^\ell \to e^{\mu \ell}$).
Finally,   Eqs[\ref{sym}-\ref{exterior}-\ref{mixed}] enable us
to decompose each irreducible summand of $\D_a \otimes \D_a \otimes \D_a$
 with respect
to the symmetry of Young tableau. For instance,
if $\D_a= \D_{\frac12}$, one obtains

\beqa
\label{D12}
\Sy^3(\D_{\frac12}) = \D_{\frac32}, \ \ 
{\$}^{^{{\tiny \begin{tabular}{|c|c|}\hline
 \ \ &  \ \ \\ \hline
 \\
\cline{1-1}  
\end{tabular}}}} \hskip -.3truecm \left({\cal D}_{\frac12}\right)=
\D_{\frac12} \oplus \D_{\frac12}.
\eeqa

\noi
and for  $\D_{\frac32}$  one gets

\beqa
\Sy^3(\D_{\frac32}) = \D_{\frac92} \oplus \D_{\frac54} \oplus \D_{\frac32}, \ \
{\$}^{^{{\tiny \begin{tabular}{|c|c|}\hline
 \ \ &  \ \ \\ \hline
 \\
\cline{1-1}  
\end{tabular}}}} \hskip -.3truecm \left({\cal D}_{\frac12}\right)= 
2 \D_{\frac72} \oplus 2 \D_{\frac52} \oplus 2 \D_{\frac32} 
\oplus 2 \D_{\f}, \ \ 
\Lambda^3(\D_{\frac32}) = \D_{\frac32}
\eeqa
\noi
Thus only a type $I_M$ mapping can be obtained with $D_\f$ 
or $\D_{\frac32}$, and the decomposition (\ref{D12}) corresponds to
(\ref{spin}). 
In fact, more generally, 
it can be proven when $a \in \f \mathbb N$ only types
$I_M$ mappings are allowed.

\subsection{Lie algebras of order three associated with the Poincar\'e
algebra}

Combining the two results established in this section (study of the finite dimensional
representations of $\iso$ and study of the $\so-$equivariant mappings from $\Sy^3(\g_1)
\longrightarrow \D_{\f,\f}$)
 non-trivial extensions of the Poincar\'e algebra 
associated to a Lie algebra of order 3, $\g=\iso \oplus \g_1$
are obtained as follow.

\begin{enumerate}
\item We consider $\g_1$ a given finite-dimensional (reducible) representation of $\so$ 
such that 
there exists an $\so-$equivariant mapping from $\Sy(\g_1)$  into $\D_{\f,\f}$ 
(of type $I$ and/or type $II$  and/or type $III$ above). 
\item  We extend the action of
$\so$ onto $\g_1$  to an  action of the  Poincar\'e algebra  $\iso$  on
$\g_1$ as in Lemma 5.1 of \cite{grt}. 
\item Assuming that $\g= \iso \oplus \g_1$ is a Lie algebra of order 3, means that some
identities have to be satisfied. These identities come from the point 1.-3. in
the Definition \ref{flie} and from the identity (\ref{J4}):

\beqa
\begin{array}{lllr}
0&=&\left[\left[X_1,X_2\right],X_3\right] +  
\left[\left[X_2,X_3\right],X_1\right] +
\left[\left[X_3,X_1\right],X_2\right],&  \text{J1}  
 \cr
0 &=&\left[\left[X_1,X_2\right],Y_3\right] +
\left[\left[X_2,Y_3\right],X_1\right] +
\left[\left[Y_3,X_1\right],X_2\right],& \text{J2}
\cr
0&=&\left[X_1,\left\{ Y_1,Y_2,,Y_3\right\}\right] -
\left\{ \left[X_1,Y_1 \right],Y_2,Y_3\right\}  -
\left\{ Y_1,\right[X_1,Y_2\left],Y_3\right\}  -
\left\{Y_1,Y_2,\left[X_1,Y_3\right] \right\},&\text{J3} 
\cr
0&=&\left[ Y_1,\left\{ Y_2,Y_3,Y_4\right\} \right]
+\left[ Y_2,\left\{ Y_3,Y_4,Y_1\right\} \right]
+\left[ Y_3,\left\{ Y_4,Y_1,Y_2\right\} \right]
+\left[ Y_4,\left\{ Y_1,Y_2,Y_3\right\} \right]
,&\text{J4}
\end{array} \nonumber
\eeqa

(for all $X_1,X_2,X_3 \in \iso, Y_1,Y_2,Y_3,Y_4 \in \g_1$).
The identities J1-J2 are satisfied by definition
 and the identity J3 by construction.
However, the identity J4 put severe constraints.
Indeed, imposing J4 automatically leads to a trivial action
of $P_\mu$ on $\g_1$ {\it i. e.} $ [P_\mu,\g_1]$=0 \cite{grt}.
(It should be noticed that if from the very beginning we have assumed that the generators
of space-time translations commute with the generators of $\g_1$, the identity J4 would
have been trivially satisfied.)  
\end{enumerate}
The point 1-3 above means that any representation $\g_1$ of $\so$  such
that $\D_{\f,\f} \subset \Sy^3(\g_1)$ leads to a possible non-trivial
extension of the Poincar\'e algebra. At that point among the three types of
brackets of a Lie algebra of order 3, 
(1) $[\iso,\iso] \subseteq \iso$, (2) $[\iso,\g_1] \subseteq \g_1$
and $\left\{\g_1,\g_1,\g_1\right\}\subseteq \D_{\f,\f}$, only the last ones 
are still unspecified. Indeed, we just know that there exists an $\so-$equivariant mapping
from $\Sy^3(\g_1)$ into $\D_{\f,\f}$.
The precise form
of these brackets, not yet known, is obtained with  identity J3.
The point 1. above ensures that the identity J3 will give non-trivial 
brackets.

 For instance, if one
start with $\g_1=\D_{\f,\f}= \left<V_\mu\right>$, we know that there
exists an $\so-$equivariant mapping $\Sy^3(\D_{\f,\f}) \to
\D_{\f,\f}$. To give the precise form of the trilinear brackets, we 
introduce  the eigenvectors of the Cartan subalgebra of $\so$:
$ (P_0,P_1,P_2,P_3) \to( P_{+,+}, P_{--,+}, P_{+,-}, 
P_{-,-})$ and similarly for $(V_\mu, \mu=0,\cdots,3)$
(see \cite{grt} for precise notations). Now a simple
weight argument gives the only non-trivial brackets :

\beqa
\label{trilin}
\begin{array}{ll}
\left\{V_{++}, V_{++},V_{--}\right\}= \alpha_1 P_{++},&
\left\{V_{--}, V_{--},V_{++}\right\}= \beta_1 P_{--}, \\
\left\{V_{++}, V_{+-},V_{-+}\right\}= \alpha_2 P_{++},&
\left\{V_{--}, V_{-+},V_{+-}\right\}= \beta_2 P_{--}, \\
\left\{V_{++}, V_{+-},V_{--}\right\}= \gamma_1 P_{+-},&
\left\{V_{--}, V_{-+},V_{++}\right\}= \delta_1 P_{-+},  \\
\left\{V_{+-}, V_{+-},V_{-+}\right\}= \gamma_2 P_{+-},&
\left\{V_{-+}, V_{-+},V_{+-}\right\}= \delta_3 P_{-+}.  
\end{array}
\eeqa

\noi
Imposing the Jacoby identity J3 we obtain 
$\alpha_2 = -\frac12 \alpha_1, \beta_1 = \alpha_1,
\beta_2 =-\frac12 \alpha_1, \gamma_1=\frac12  \alpha_1,
\gamma_2 = - \alpha_1,
\delta_1=\frac12  \alpha_1,\delta_2 = - \alpha_1,$ 
and the brackets (\ref{trilin}) reproduce the trilinear brackets of Example
\ref{FP} : 
$\left\{V_\mu, V_\nu,V_\rho\right\} = \eta_{\mu \nu} P_\rho + \eta_{\nu \rho} P_\mu +
\eta_{\rho \mu} P_\nu$ \cite{grt}. Thus,
 the only Lie algebra of order 3 associated to
$\g=\iso \oplus \D_{\f,\f}$ is the Lie algebra of order 3 given in  Example \ref{FP}. \\

This general study shows that many Lie algebras of order 3 extending non-trivially the
$(1+3)-$dimensional Poincar\'e algebra can be defined.

\bigskip

{\bf Acknowledgment:} P. Baumann is gratefully acknowledge for useful discussions
and remarks. The Organizing Committee, and in particular 
G. Pogosyan are kindly acknowledged for the friendly and
studious atmosphere during the conference.

\end{document}